\documentclass[twocolumn]{svjour3}         
\smartqed  
\usepackage{graphicx}
\newcommand{\bce}{\begin{center}}
\newcommand{\ece}{\end{center}}
\newcommand{\be}{\begin{equation}}
\newcommand{\ee}{\end{equation}}
\newcommand{\bea}{\begin{eqnarray}}
\newcommand{\eea}{\end{eqnarray}}
\newcommand{\E}{\> = \>}
\newcommand{\EA}{&=&}
\newcommand{\non}{\nonumber\\}
\newcommand{\To}{\> \longrightarrow \>}
\newcommand{\vecxi}{\mbox{\boldmath$\xi$}}

\newcommand{\fb}{{\bf b}}  
\newcommand{\fk}{{\bf k}}
\newcommand{\fK}{{\bf K}}
\newcommand{\fp}{{\bf p}}
\newcommand{\fq}{{\bf q}}
\newcommand{\fr}{{\bf r}}
\newcommand{\fv}{{\bf v}}
\newcommand{\fw}{{\bf w}}
\newcommand{\fu}{{\bf u}}
\newcommand{\fx}{{\bf x}}
\newcommand{\fy}{{\bf y}}
\newcommand{\Tint}{\int_{-T}^{+T} dt}
\newcommand{\Def}{\> := \>}

\journalname{Few-Body Systems}
\begin{document}

\title{Exact Path-Integral Representations for the $T$-Matrix in Nonrelativistic 
Potential Scattering
\thanks{Contribution to the workshop ``Relativistic Description 
of Two- and Three-Body Systems in Nuclear Physics'', ETC*, October 19-23, 2009}
}

\titlerunning{Path integrals for the $T$-matrix}   

\author{R. Rosenfelder}


\institute{R. Rosenfelder \at
              Paul-Scherrer-Institut, CH-5232 Villigen PSI, Switzerland \\
              Tel.: +41-56-310-3663\\
              \email{roland.rosenfelder@psi.ch}           
}

\date{Received: 29 July 2010 / Accepted: 16 September 2010 /
Published in Few-Body Syst. {\bf 49}, 41 -- 50 (2011)}

\maketitle

\begin{abstract}
Several path integral representations for the $T$-matrix 
in nonrelativistic potential scattering are given which produce
the complete Born series when expanded to all orders and the eikonal
approximation if the quantum fluctuations are suppressed.
They are obtained with the help of ``phantom'' degrees of freedom which 
take away explicit phases that diverge for asymptotic times. 
Energy conservation is enforced by imposing a Faddeev-Popov-like constraint 
in the velocity path integral. 
An attempt is made to evaluate stochastically the real-time path integral 
for potential scattering and generalizations to relativistic scattering are discussed.

\end{abstract}

\section{Introduction}
\label{intro}
Nonrelativistic quantum mechanical scattering in a local potential 
is usually described
in the framework of time-dependent or time-independent solutions of the
Schr\"odinger equation.
Path-integral methods in quantum mechanics, on the other hand,
are mostly applied to the discrete spectrum, e.g. for harmonic 
or anharmonic oscillators.
In contrast, the transition matrix for the continuous spectrum is rarely 
represented as path integral. Even if available, many  
representations turn out to be rather formal, e.g. requiring
infinitely many differentiations or infinite time limits
to be performed. This is not only impractible but also 
unfortunate since a convenient path integral representation 
may lead to new approximations and may be extended readily to 
the many-body problem or Quantum Field Theory. Also the long-standing problem 
how to evaluate 
{\it real-time} path integrals by stochastic methods needs a suitable 
path integral representation as starting method. There has been significant 
progress in dealing with real-time path integrals for dissipative 
systems  but in closed systems and infinite scattering 
times only zero-energy scattering seems to be tractable by Euclidean Monte-Carlo
methods at present \cite{KML}. 

Here I will describe an approach developed recently \cite{Rose} to overcome 
these difficulties while still being practical. Its main features are the
use of ``phantom'' degrees of freedom to cancel phases which would diverge for 
asymptotic scattering times. The eikonal approximation - valid for high energy 
and small scattering angles and the basis for Glauber's multiple scattering
approach - is ``built in'' from the beginning as it is obtained when 
all quantum fluctuations are neglected.

\section{How to obtain a path-integral representation of the $T$-matrix}
\label{sec:1}
We start with the definition of the $S$-matrix as matrix element of the 
time-evolution operator in the interaction picture at infinite scattering times
\bea
{\cal S}_{i \to f} \EA \lim_{T \to \infty} \> \left < \phi_f \left | \,
\hat U_I (T,-T) \, \right | \phi_i \right > \non
\EA \lim_{T \to \infty} \>
e^{i (E_i + E_f) T} \, \left < \phi_f \left | \,
\hat U(T,-T) \, \right | \phi_i \right >
\label{S-Matrix}\non
&=:&
(2 \pi)^3 \delta^{(3)} \left ( \fk_i - \fk_f \right )  - 2 \pi i
\delta\left ( E_i - E_f \right ) \, {\cal T}_{i \to f}
\eea
and its connection with the $T$-matrix.
Here the scattering states are normalized according to 
$ \> \left < \phi_f | \phi_i \right >
= (2 \pi)^3 \delta^{(3)}( \fk_f - \fk_i ) \> $.
With the definition of $\hat U_I$ in terms of the full time-evolution operator
$\hat U$ one could use its standard representation as a path integral over
trajectories ($\hbar = 1$  )
\bea
U(\fx_b, T; \fx_a, -T) \EA \int_{\fx(-T)=\fx_a}^{\fx(T)=\fx_b} {\cal D}^3 x(t)
\non
&& \hspace{-2cm}  \times \,  \exp \left [ \, i \Tint  \left ( \frac{m}{2} 
\dot \fx(t)^2 - 
V(\fx(t)) \right )\, \right ] 
\label{Lagrange PI}
\eea
connecting initial and final points.
However, keeping the boundary conditions of the paths is cumbersome and it is 
more convenient to convert the path integral into one over {\it velocities} by
inserting \footnote{Better done in the discretized version of the path integral
which shows that one has  $N \> \fv$-integrations compared to $N-1$ ones 
over the intermediate points of the trajectory when the endpoints are fixed.}
\be
1 \E \int {\cal D}^3 v \> \delta^{(3)} \left ( \fv(t) - \dot \fx(t) \right )
\> .
\ee
Then one obtains
\bea
\left ( {\cal S} - 1 \right )_{i \to f} \EA \lim_{T \to \infty} 
e^{i \Phi(T) } \> \int d^3 r \>
e^{- i \fq \cdot \fr} \non
&& \hspace{-0.5cm} \times  \int {\cal D}^3 v \> \exp \left [ \,
i \Tint \, \frac{m}{2} \fv^2(t) \, \right ]   \non
&& \hspace{-2cm} \times \left \{ \, \exp \left [  - i \Tint \,
V \left (  \fr + \frac{\fK}{m} t + \fx_v(t)
\right ) \right ] - 1 \, \right \} \> ,
\label{S PI}
\eea
with 
\be
\fx_v(t) \E \frac{1}{2} \Tint' \> {\rm sgn}(t-t') \, \fv(t') , 
\quad \dot \fx_v (t) \E v(t) \>. 
\label{connection x,v}
\ee
Here 
\be
\fq \E \fk_f - \fk_i \> , \quad \fK \E \frac{\fk_i + \fk_f}{2}
\ee
and the path integral is normalized such that
\be
\int {\cal D}^3 v \> \exp \left [ \,
i \Tint \, \frac{m}{2} \, \fv^2(t) \, \right ]   \E 1 \> .
\label{norm}
\ee
There remain two problems to obtain the desired path-integral representation of 
the $T$-matrix:
\begin{description}
\item[1.] There is still an explicit phase
\be
\Phi(T) \E \left (E_i + E_f - \frac{\fK^2}{m} \right ) \, T \> \equiv \> 
\frac{\fq^2}{4 m} \, T
\ee
which seems to produce a divergence in the limit $T \to \infty$.
Of course, it is cancelled in each order of perturbation theory as one can check.

\item[2.] Energy conservation is not yet evident.

\end{description}
\vspace{0.2cm} 

The first problem is solved by realizing that this phase factor is generated by 
applying $ \exp ( - i T \Delta/(4m) )$ on $\exp ( - i \fq \cdot \fr) $. 
Integrating by parts and ``undoing the square'' by means of a path integral
over an ``antivelocity''
\bea
\exp \left ( -\frac{i}{4 m} T \, \Delta \right ) \EA 
\int {\cal D}^3 w \,
\exp \Bigl [ - i \Tint  \, \frac{m}{2} \, \fw^2(t) \non
&& \pm \Tint \> \frac{1}{2} \, f(t) \,
\fw(t) \cdot \nabla \> \Bigr ] 
\label{undoing}
\eea
this simply becomes a shift operator. In Eq. (\ref{undoing}) the arbitrary real
function only has to fulfill $ \Tint f^2(t) = 2 T $ and we choose it 
for simplicity as $ f(t) = {\rm sgn}(t)$. Note that the sign of the 
kinetic energy for the antivelocity is opposite to the usual kinetic energy
in order to obtain a {\it real} shift in the argument of the potential. This has 
an amusing similarity with the ``phantom'' degrees of freedom in the Lee-Wick 
approach to quantum field theory \cite{LeeWick}
which has been discussed again recently 
\cite{GCW}.

Without a perturbative expansion of the $S$-matrix the second problem 
is solved by the Faddeev-Popov trick of multiplying the path integral with
\be
1 \E \frac{|\fK|}{m} \int_{-\infty}^{+\infty} d\tau \>
\delta \left ( \> \hat
\fK \cdot \left [ \fr + \frac{\fK}{m} \tau \right ] 
\> \right ) \> ,
\label{FP1}
\ee
and shifting the co-ordinates. Arguing that in the limit $T \to \infty $ 
the action is invariant under a finite time shift $\tau$ one obtains
\bea
\int d^3r \> e^{-i \fq \cdot \fr} \> \ldots &\To&
\frac{|\fK|}{m} \int_{-\infty}^{+\infty} d\tau
\exp \left ( - i \fq \cdot \frac{\fK}{m} \tau \right ) \non
&& \times \int d^3r \> \delta \left (
\hat \fK \cdot \fr \right ) e^{- i \fq \cdot \fr} \> \ldots \non
\EA 2 \pi \, \delta \left ( \frac{\fq \cdot \fK}{m} \right ) \> \int d^2 b 
\> e^{- i \fq \cdot \fb}
\> \ldots \non
&& \hspace{-1.5cm} \E 2 \pi \delta \left ( \frac{\fk_f^2}{2m} - 
\frac{\fk_i^2}{2m} \right ) \>
\int d^2 b \> e^{- i \fq \cdot \fb} \> \ldots 
\eea
Actually, this procedure is a delicate exchange of time-limits but it
has been checked that the outcome is correct by deriving the Born approximation
to all orders (see appendix of Ref. \cite{Rose}).
So we obtain the following path-integral representation of the $T$-matrix
\bea
{\cal T}_{i \to f}^{(3-3)} \EA i \frac{K}{m}
\> \int d^2 b \> e^{- i \fq \cdot \fb } \>
 \int {\cal D}^3 v \, {\cal D}^3 w \non 
&& \times \exp \left \{  \,
i  \, \int_{-\infty}^{+\infty} dt \>  \frac{m}{2} \,
 \biggl [ \, \fv^2(t) - \fw^2(t) \, \biggr ]\, \right \} \non
&& \quad \times \Bigl \{ \>  \exp \left [ \> i \, \chi(\fb,\fv(t),\fw(t)) \> \right ]- \, 1 
\> \Bigr \}
\label{3-3}
\eea
with
\bea 
\chi(\fb,\fv,\fw)  \EA -  \int_{-\infty}^{+\infty}
dt \> V ( \vecxi(t) ) \non 
\vecxi(t) \EA \fb + \frac{\fK}{m} \, t  + \fx_v(t) - \fx_w(0) \> .
\label{def xi}
\eea
Note that $\dot{\vecxi}(t) = \fK/m + \fv(t)$ and 
\be
|\fK| \E k \cos (\theta/2) \> , \> \> |\fq| \E 2 k \sin (\theta/2 )
\ee
where $\theta$ is the scattering angle and $E = k^2/(2m)$ the scattering energy.
Eqs. (\ref{3-3}) and (\ref{def xi}) show 
that the particle travels along a straight-line reference trajectory 
parallel to the mean momentum
and that the  quantum fluctuations are taken into account by the 
functional integral over velocity and antivelocity.

\begin{figure}
\hspace*{0.7cm}\includegraphics[width=0.85\linewidth]{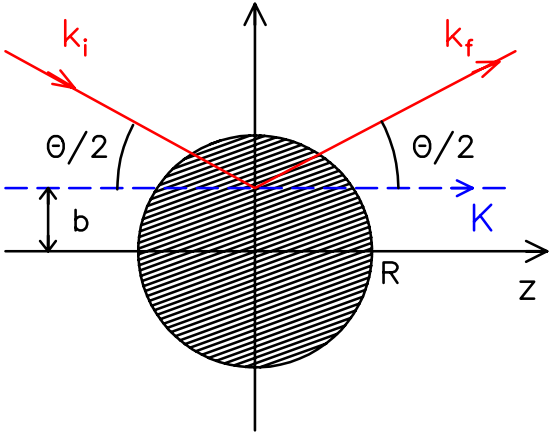}
\caption{(Color online). Scattering geometry for a potential of radius $R$ showing
the impact parameter $b$, the ray made by the incoming and outgoing momenta 
$ \> {\bf k}_{i,f} \> $, and
the mean momentum $ \fK = (\fk_i + \fk_f)/2 $.
}
\label{fig:1}
\end{figure}

In Eq. (\ref{3-3}) a 3-dimensional antivelocity is used as indicated by the
superscript ``(3-3)''. By a simultaneous shift of velocity 
variables and impact parameter $\fb$ it is possible to obtain a representation
utilizing only a 1-dimensional longitudinal 
(i.e. parallel to the mean momentum $\fK$) antivelocity:
\bea
{\cal T}_{i \to f}^{(3-1)} \EA i \frac{K}{m} \> \int d^2 b \>
e^{- i \fq \cdot \fb}
\>  \int {\cal D}^3 v  \, {\cal D}w \non
&& \times \exp \left \{ \, i  \,  \int_{-\infty}^{+\infty} dt \>  
\frac{m}{2} \, \biggl [ \, 
\fv^2(t) - w^2(t) \, \biggr ] \, \right \} \non
&& \times \Biggl \{  \>  \exp \left [ \, - i \, \int_{-\infty}^{+\infty} dt
\>  V \left ( \vecxi_{\rm ray}(t) \right ) \right ]  - 1 \> \Biggr \} 
\eea
with
\bea
\vecxi_{\rm ray}(t) \EA \fb + \frac{\fp_{\rm ray}(t)}{m} t +
\fx_v(t) - \fx_{v \, \perp}(0) - x_w(0)  \hat \fK \non
 \fp_{\rm ray}(t) \EA \fK + \frac{\fq}{2} \, {\rm sgn}(t) \E
{\bf k}_i \, \Theta(-t) + {\bf k}_f \, \Theta(t) \> .
\eea
Here the particle travels along a {\it ray} made by the incident momentum 
for $t < 0 $ and the final momentum for $t > 0$. This is depicted in 
Fig. \ref{fig:1}

What happens for a {\it nonlocal} potential $\hat V$? In this case we have to 
start not from the Lagrangian path integral (\ref{Lagrange PI}) but from the 
more general phase-space path integral
\bea
U(\fx_b, T; \fx_a, -T) \EA \int_{\fx(-T)=\fx_a}^{\fx(T)=\fx_b} {\cal D}^3 x(t)\, 
\int{\cal D}^3 p(t)\non
&& \hspace{-1cm}  \times \,  \exp \Biggl \{ \, i \Tint  \Bigl [ \fp(t) \cdot \dot 
\fx(t) - \frac{\fp^2(t)}{2m} \non 
&&  \hspace{1cm} - V_W(\fx(t),\fp(t)) \Bigr ]\, \Biggr \} \>.
\label{phase space PI}
\eea
Here
\be
V_W(\fx,\fp) \E \int d^3y \> \left < \fx - \frac{y}{2} | \hat V | \fx 
+ \frac{y}{2} \right > \,e^{i \fp \cdot \fy}
\label{Wigner trans}
\ee
is the Wigner transform of the quantum-mechanical operator $\hat V$. Using it in the 
phase-space path integral is equivalent to applying 
Weyl's quantization rule to velocity-dependent interactions
(the ``mid-point'' rule). One can now follow all the previous steps to derive a 
 path-integral representation of the $T$-matrix. Writing the additional 
(unconstrained) momentum
variable $ \fp(t) = m \, \fu(t) $ and using a 3-dimensional antivelocity 
we then arrive at the following expression
\bea
{\cal T}_{i \to f}^{(3-3)} \EA i \frac{K}{m}
\> \int d^2 b \> e^{- i \fq \cdot \fb } \>
 \int {\cal D}^3 v \, {\cal D}^3 w \,  {\cal D}^3 u \non 
&& \times \exp \left \{ \,
i  \, \int\limits_{-\infty}^{+\infty} dt \,  \frac{m}{2} \,
 \biggl [ \, \fv^2(t) - \fw^2(t) - \fu^2(t) \, \biggr ]\,
\right \} \non
&& \times  \Bigl \{ \>  \exp \left [ \> i \,\chi(\fb,\fv,\fw,\fu) \> \right ] - 1 \> 
\Bigr \}
\label{3-3 nonlocal}
\eea
where the phase function(al) is now
\be
\chi(\fb,\fv,\fw,\fu)=- \! \int\limits_{-\infty}^{+\infty}
dt \, V_W \bigl [  \vecxi(t), \fK + m \fv(t) + m \fu(t) \bigr]
\ee
and $\vecxi(t)$ has been defined in Eq. (\ref{def xi}). Again the path integrals 
are normalized such
that Gaussian integrals are unity as in Eq. (\ref{norm}).

\section{Applications}
\subsection{Eikonal expansions}

For high-energy and small scattering angles the particle mainly travels along 
the reference
path as shown by an appropriate scaling of variables \cite{Rose}: all quantum 
fluctuations 
are suppressed by inverse powers of $K = k \cos(\theta/2)$ for the representation 
with a 3-dimensional antivelocity and by inverse powers of $k$ for the ``ray'' version.
In the former case the leading term is given by 
\be
{\cal T}_{i \to f}  \!\simeq \!\frac{iK}{m} \!
\int  \! \! d^2 b  e^{- i \fq \cdot \fb}  \Biggl \{   
\exp \left [-i   \int\limits_{-\infty}^{+\infty} \! dt  V \left (\fb+
\frac{\fK t}{m} \right ) \right ] - 1 \Biggr \},
\label{AI 0}
\ee
a version of the eikonal approximation due to Abarbanel and Itzykson \cite{AI}. 
Higher-order
corrections can be calculated systematically and agree with the systematic 
eikonal expansion due
to Wallace \cite{Wal}. In the nonlocal case neglecting all quantum fluctuations 
$\fv(t), \fw(t), 
\fu(t) $ in Eq. (\ref{3-3 nonlocal}) gives an approximation for the $T$-matrix 
first derived in Ref. \cite{Rowig}.

\subsection{Variational approximations}

The new path-integral representations for the $T$-matrix suggest new variational 
approximations by applying the Feynman-Jensen variational principle. 
This has been studied in Ref. \cite{Carr}
with a trial action which is linear in the velocities $\fv(t), \fw(t) $ and in 
Ref. \cite{CaRo} with a more general (linear + quadratic) trial action. 
In both cases the first correction to the variational result from a cumulant 
expansion has also been evaluated
and impressive agreement with partial-wave calculations has been found, even for larger 
scattering angles.

\subsection{Stochastic evaluation of high-energy scattering}

One may try to evaluate the path integral for the $T$-matrix numerically by
stochastic methods as is done in the Euclidean case. For this purpose 
(and for simplicity in the representation with a 3-dimensional antivelocity) we 
expand the velocity variables in a complete set of harmonic oscillator functions
\be
\left ( \begin{array}{c}
          \fv(t) \\
          \fw(t) \end{array} \right ) \E  C \, \sum_{n=0}^{\infty} \, 
\left ( \begin{array}{c}
          \fv_n \\
          \fw_n \end{array} \right ) \, u_n \left ( \frac{t}{t_0} \right ) \> .
\label{v,w ho expan}
\ee
Here $t_0$ denotes a characteristic time 
which we take as the one needed to traverse the 
range $R$ of the potential, i. e. $t_0 = mR/k$ and the constant $C$ is chosen 
such that the free action becomes 
\be
S_0 \E \sum_{n=0}^{\infty} \left ( \fv_n^2 - \fw_n^2 \right ) \> .
\label{S0}
\ee
Writing
\be
\vecxi(t) \E \fb + \frac{\fK}{m} t + \fx_{\rm quant}(t)
\label{def x_quant}
\ee
the quantum trajectory is then given by 
\be 
\fx_{\rm quant}(t) \E \sum_{n=0}^{\infty} \left [ \, q_n \left ( \frac{t}{t_0} 
\right )  \> \fv_n - q_n(0) \, \fw_n \, \right ]
\ee
where 
\be
q_n(y) 
\Def \frac{C t_0 }{2} \, \int_{-\infty}^{+\infty} dx \> {\rm sgn}(y - x) \, u_n(x)
\> .
\ee
Completeness of the harmonic-oscillator wave functions implies
\be
\sum_{n=0}^{\infty} q_n  \,  \left ( \frac{t_1}{t_0} \right ) 
q_n \left ( \frac{t_2}{t_0} \right )  
\E \frac{1}{m} \, \Bigl  [ \, T - \left | t_1 - t_2 \right | \, \Bigr ]
\label{complete}
\ee
where the divergence of the infinite sum has been regulated by a large 
scattering time $T$ as before.

The path integral (\ref{3-3}) can now be written as an infinite-dimensional integral
over the expansion coefficients
\bea
{\cal T}_{i \to f}^{(3-3)} \! \EA \! \frac{iK}{m}
 \int \! d^2 b \, e^{- i \fq \cdot \fb } \,
\left \{ \prod_{n=0}^{\infty} \! \left ( \int d^3 v_n  d^3 w_n \right ) \, 
e^{iS_0} \right \}^{-1} \non
&& \quad \times \prod_{n=0}^{\infty} \left ( \int d^3 v_n \, d^3 w_n \right )\> e^{i S_0} \, 
\Bigl ( e^{ i\chi} \, - \, 1 \> \Bigr ) \> . 
\label{3-3 in modes}
\eea
In any numerical calculation the infinite sum over the modes $n$ has 
to be reduced to a finite
one involving only $N$ modes. 
Therefore we split the free action into two parts , one ($S_0^<$) involving 
the lower modes $n = 0, \, \ldots N-1$ and the other ($S_0^>$) depending only on the
``upper'' expansion coefficients
$n = N, \, \ldots \infty$. 

For the upper modes we employ the method of ``partial averaging'' \cite{PA}, i. e.
\bea
\left < \, e^{i \chi} \, \right >_N^{\infty} \> 
&:=& \frac{\prod_N^{\infty} \left ( \int d^3 v_n d^3 w_n \right ) \> 
\exp \left ( \, i S_0^{>} \, \right ) \> e^{i \chi}}{ \prod_N^{\infty} \left (\int
 d^3 v_n d^3 w_n \right ) \> \exp \left ( \, i S_0^{>} \, \right ) } \non
&\simeq& \exp \left ( \> i \left < \, \chi \, \right >_N^{\infty} \> \right ) \> .
\label{PA}
\eea
In Fourier path integrals
partial averaging has been successfully applied to determine
the ground-state and thermal
properties of bound systems (see, e.g. Ref. \cite{AlRo}). It includes
the upper modes (up to infinity) at least approximately by averaging them 
with the free action thereby improving convergence with the number $N$ of explicit modes 
\cite{KFD}.

Here we apply this method to a scattering problem.
Note that $ \left < \chi \right >_N^{\infty} $ depends on
 the lower coefficients and acts as a (complex) interaction phase for the 
$6N$-dimen\-sional integral over the coefficients $\fv_n, \fw_n , \>
n = 0, \, \ldots N-1 $.
The free average of the potential over the upper coefficients is readily performed as only 
Gaussian integrals have to be evaluated and one obtains
\be
\left < \chi \right >_N^{\infty}  \E - \int_{-\infty}^{+\infty}dt \>
V_{\sigma(t)} \left ( \fx_{\rm quant}^{<}(t) \right )
\ee
where
\be
\fx_{\rm quant}^{<}(t) \Def \sum_{n=0}^{N-1} \left [ \, q_n \left ( \frac{t}{t_0} 
\right ) \, \fv_n -  q_n (0) \, \fw_n \, \right ]
\ee
is the quantum trajectory involving the lower coefficients and 
\be
\tilde V_{\sigma(t)}(\fp) \Def \tilde V(\fp) \,\exp \left [\,  - \frac{1}{2} \,
\sigma(t) \fp^2 \, \right ] \> ,
\label{def Vsigma}
\ee
the Gaussian transform of the potential in momentum space. The width 
turns out to be purely imaginary
\be
\sigma_N(t/t_0) \Def \frac{i}{2} \sum_{n=N}^{\infty} \left [ \, 
q_n^2 \left ( \frac{t}{t_0} \right ) -  q_n^2(0) \, \right ] 
\ee
where the first term in the bracket comes from the velocity integration and the last one
from the integration over the antivelocity.
Due to Eq. (\ref{complete}) we can write it as
\be
\sigma_N(t/t_0) \E \frac{i}{2 m} \left [ \, T -   T \, \right ] +  
\frac{i}{2} \sum_{n=0}^{N-1} 
\left [ \, q_n^2(0) -  \, q_n^2 \left ( \frac{t}{t_0} 
\right ) \, \right ] 
\label{cancell in sigma}
\ee
showing how the arising divergence is exactly cancelled
by the contribution of the antivelocity.

Still it is impossible to evaluate the remaining $6N$-dimensional integral 
over the lower coefficients without any damping.
This is provided by giving the particle a complex mass \footnote{Actually, this is the
method chosen in Ref. \cite{JoLa} for rigorously defining real-time Feynman path integrals.}
\be
m_{\rm particle} \>  \longrightarrow \> m \, \left ( 1 + i \Gamma \right ) 
\ee
and the phantom the mass $ m^{\star}$ 
\be
m_{\rm phantom}  \longrightarrow  m \, \left ( 1 - i \Gamma \right )  \> . 
\ee
If this is done in the full path integral, $\Gamma \to \infty$ would lead to
the eikonal approximation because all quantum fluctuations are suppressed in that limit
and $\Gamma \to 0$ would give the exact result. This supports the hope 
that numerical evaluations
at finite $\Gamma$ and an extrapolation of these results  to $\Gamma \to 0 $ 
would yield reasonably accurate results for the $T$-matrix. 

However, it is not possible 
to modify the path integral over the upper coefficients in this way
as the phantom must have the same mass as the particle:  If it would have a 
different mass, say $\lambda m $ then the second term on the r.h.s of 
Eq. (\ref{cancell in sigma}) 
would be multiplied by $1/\lambda$ and the diverging term $T$ (originating from the particle)
would not be cancelled. So partial averaging and damping do not marry (easily) ... 

\begin{figure}
\hspace*{0.7cm}\includegraphics[width=0.85\linewidth]{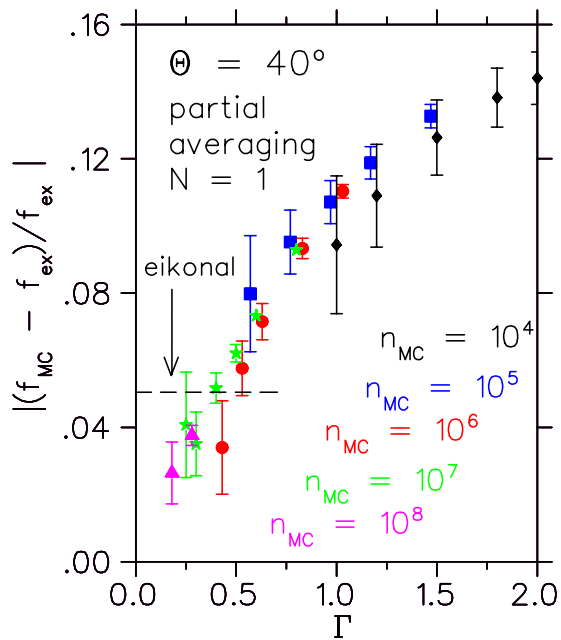}
\caption{(Color online). The relative deviation $|\Delta f/f |$ of the stochastic 
scattering amplitude 
from the exact partial wave result at a fixed scattering angle $\theta = 40^o$ for scattering
from the  Gaussian potential (\ref{Gauss pot}) with 
$ 2m V_0 R^2 = -4$ at $k R = 4 $.
}
\label{fig:2}
\end{figure}

If they remain separated, i. e. if only the explicit modes are damped then
the stochastic results only slowly converge to the exact ones if $\Gamma$ is decreased. 
This is shown in Fig. \ref{fig:2} 
where scattering from a Gaussian potential 
\be
V(r) \E V_0 \, e^{-r^2/R^2} 
\label{Gauss pot}
\ee
at a fixed scattering angle is considered for $ N = 1 $. It is seen that the relative 
deviation of the stochastic scattering amplitude indeed decreases 
if the damping is made smaller 
but that one also needs more and more Monte-Carlo calls 
to obtain a statistically valid result.

Although encouraging these results indicate that further work is needed 
to obtain a practical Monte-Carlo scheme for high-energy potential scattering. 
For a different approach see Ref. \cite{Smir}.

\section{Relativistic extensions}

To stay closer to the spirit and the aims of this workshop I will now consider scattering 
under relativistic conditions, i.e. truly high-energy scattering. In my opinion
quantum field theory provides the most
general, consistent - albeit not the most practical - framework to describe these
phenomena and therefore I will consider a scalar theory of particles ($\Phi$) with mass $M$
interacting by exchange of scalar particles ($\phi$) of mass $m$ given by an interaction 
Lagrangian
\be
{\cal L}_I \E g \, \left |\Phi^2 \right | \, \phi \> .
\ee
This generalized Wick-Cutkosky model \cite{Wick,Cut} mimics meson-exchange 
between massive nucleons and
-- 55 years after its introduction --  is still the favorite model~ 
\footnote{Unfortunately, it also has an unstable ground state which is 
ignored by many authors, even 50 years after this unpleasant fact has been 
proven \cite{Baym}. For the present discussion I will join this group ...}
of many workers in the field 
mostly for the relativistic bound-state problem (see, e.g. Ref. \cite{BRS}).

Let us evaluate the 4-point function of the theory: neglecting nucleon loops (``quenched 
approximation'') and using the Schwinger representation for the nucleon propagator
in the presence of the meson field
\bea
&&\left [-\partial^2 - M^2 + g \, \phi(x) + i0 \right ]^{-1} \E \non 
&& \quad \frac{1}{i M} \int_0^{\infty} dT \> \exp \left [  \frac{i}{M} 
\left ( -\partial^2 - M^2 + g \, \phi(x) \right ) T 
\right] 
\label{Schwinger trick}
\eea 
the 4-point function for scattering of two nucleons can be written in ``worldline'' form
\bea
&& G_4(x_1,x_2,x_3,x_4) =  \int_0^{\infty}\! \! dT_1 \, dT_2 \, \exp \left [  
- i M \left ( T_1 + T_2 \right ) \right ] \non
&& \times \, \int_{y_1(-T_1) = x_1}^{y_1(+T_1) = x_2}
{\cal D}^4y_1  \int_{y_2(-T_2) = x_3}^{y_2(+T_2) = x_4}{\cal D}^4y_2 \non
&& \times \exp \Biggl ( \, i 
S_0 [ y_1,y_2 ] + i S_{\rm int} [ y_1, y_2 ] \, \Biggr )  
 + \left ( \, x_1 \leftrightarrow x_3 \, \right ).
\label{G4}
\eea
Here the free action is given by
\be
 S_0 [ y_1,y_2 ] \E \sum_{i=1}^{2} \int_{-T_i}^{+T_i} d\tau \left ( - \frac{M}{2} \dot y_i^2 
\right )
\ee
while the interaction part is retarded because the mesons have been integrated out:
\bea  
 S_{\rm int} [ y_1, y_2] \EA - \frac{g^2}{8 M^2}\sum_{i,j = 1}^{2}
\int_{-T_i}^{+T_i} d\tau_1 \int_{-T_j}^{+T_j} d\tau_2 \> \int \frac{d^4 p}{(2 \pi)^4} \non 
&& \hspace{-1.2cm} \times \frac{1}{p^2 - m^2 + i0} \, 
\exp \left [ \, - i p \cdot \left ( y_i(\tau_1) - y_j(\tau_2) \right ) \right ] \> .
\label{S int}
\eea
This looks (superficially) very similar to a nonrelativistic description: 
if we neglect the $(i = j)$-terms, i. e. the self-energy of the nucleons, then we have a 
4-dimensional analogue of scattering from a Yukawa potential. Still 
there are two different proper times $T_1, T_2$ over which one has to integrate 
finally and a mass renormalization is needed to get rid of the divergencies at small
proper time -- of course, all these features are to be expected in a relativistic 
quantum field theory \footnote{The Wick-Cutkosky model belongs to the simpler
class of super-renormalizable theories and is, of course, a far cry from the 
field theory of strong interactions, QCD. One could paraphrase a remark made by A. Einstein 
in 1921: "As far as the models refer to reality, they are not manageable;
and as far as they are manageable, they do not refer to reality".}.

To obtain the relativistic $T$-matrix one has to Fourier transform the 4-point function
(\ref{G4}) (easily done by converting to velocity path integrals), 
extract the energy-momentum-conserving $\delta$-function and remove the 
outer legs, i.e. multiply with the inverse (full) 2-point functions of the external
particles. While the extraction of energy-momentum conservation is obvious
in this field-theoretic framework due to the translation invariance of the action, 
the amputation of the {\it full} external nucleon propagators is more involved. 
Using the Bloch-Nordsieck approximation 
it is possible to achieve that by functional differentiation
techniques (see, e.g. Ch. 10.1 in Ref. \cite{Fried}) 
but I expect that introducing phantom degrees of freedom as in the nonrelativistic 
case will do a better job. This is presently under investigation.

\section{Summary}

Several path integral representations for the $T$-matrix in potential scattering 
have been  discussed, including a new one for nonlocal potentials. They serve 
to derive high-energy
expansions, variational approximations and are a natural starting point for a stochastic
evaluation of high-energy potential scattering. Many avenues lie open for further development,
in particular in the relativistic domain.


%
%

\begin{acknowledgements} I would like to thank the organizers of this workshop, 
in particular Giovanni Salm\`e, for an inspiring and fruitful meeting. 
Many results of this work
were obtained in an enjoyable collaboration with Julien Carron to whom I am very much 
indebted. 
\end{acknowledgements}

\vspace{0.3cm}

\renewcommand{\baselinestretch}{1.}

\footnotesize
\baselineskip10pt
\noindent
{\bf Note added after publication:} $\>$ A method for non-perturbative amputation of the 4-point function in 
the Wick-Cutkosky model and in scalar QED has been 
developed by Barbashov and colla\-borators \cite{Barb1,Barb2,MaTa}. See also Refs. \cite{HaXu,HaPo}.
\normalsize


\end{document}